\documentclass[preprint,12pt,epsfig,axocolor,epsf]{article}
\pdfoutput=1 

\usepackage{jheppub} 
\usepackage{amsmath}
\usepackage{esvect}
\usepackage{amstext}
\usepackage{amssymb}
\usepackage{lineno,hyperref}
\usepackage{makecell}
\usepackage{color}
\usepackage{soul}
\usepackage[utf8]{inputenc}
\usepackage{braket}

\usepackage[normalem]{ulem}

\definecolor{darkgreen}{cmyk}{1,0,1,0.4}
\definecolor{brown}{cmyk}{0,0.8,1,0.2}
\definecolor{darkred}{cmyk}{0,1,1,0.2}
\def\hlpm#1{\textcolor{black}{\textrm{#1}}}

\newcommand {\ignore}[1]{}

\newcommand{\be}{\begin{equation}}
\newcommand{\ee}{\end{equation}}
\newcommand{\bea}{\begin{eqnarray}}
\newcommand{\eea}{\end{eqnarray}}

\newcommand{\ldm}{\ensuremath{\Delta m_{31}^2}}
\newcommand{\sdm}{\ensuremath{\Delta m_{21}^2}}

\newcommand{\eet}{\ensuremath{\varepsilon_{e\tau}}}
\newcommand{\emt}{\ensuremath{\varepsilon_{\mu\tau}}}
\newcommand{\ett}{\ensuremath{\varepsilon_{\tau\tau}}}
\newcommand{\eee}{\ensuremath{\varepsilon_{ee}}}
\newcommand{\eem}{\ensuremath{\varepsilon_{e\mu}}}
\newcommand{\emm}{\ensuremath{\varepsilon_{\mu\mu}}}
\newcommand{\eetp}{\ensuremath{\varphi_{e\tau}}}

\newcommand{\eemp}{\ensuremath{\varphi_{e\mu}}}

\newcommand{\epst}[2]{\ensuremath{\tilde{\varepsilon}_{{#1}{#2}}}}

\begin{document}
\hspace{10pt}
\title{Non-standard neutrino oscillations:  perspective from unitarity triangles}

\author[a]{Mehedi Masud}

\author[b]{\!\!,~Poonam Mehta}

\author[c]{\!\!,~Christoph A. Ternes}

\author[d]{\!\! and~Mariam T{\'o}rtola}

\affiliation[a]{Center for Theoretical Physics of the Universe, Institute for Basic Science (IBS),\\
Daejeon 34126, South Korea}
\affiliation[b]{School of Physical Sciences, Jawaharlal Nehru University, New Delhi 110067, India}
\affiliation[c]{INFN, Sezione di Torino, Via P. Giuria 1, I--10125 Torino, Italy}
\affiliation[d]{Departament de Física Teòrica, Universitat de València, and Instituto de F\'{i}sica Corpuscular, CSIC-Universitat de Val\`{e}ncia, 46980 Paterna, Spain}

\emailAdd{masud@ibs.re.kr}
\emailAdd{pm@jnu.ac.in}
\emailAdd{ternes@to.infn.it}
\emailAdd{mariam@ific.uv.es}
 
\abstract{
{{We formulate an alternative approach based on unitarity triangles to describe neutrino oscillations in presence of non-standard interactions (NSI). 
Using perturbation theory, we derive the expression for the oscillation probability in case of NSI and cast it in terms of the three independent parameters of the leptonic unitarity triangle (LUT).  The form invariance of the probability expression (even in presence of new physics scenario  as long as the mixing matrix is unitary) facilitates a neat geometric view of neutrino oscillations in terms of LUT.  We examine the regime of validity of perturbative expansions in the NSI case and make comparisons with approximate expressions existing in literature.   We  uncover some interesting  dependencies on NSI terms while studying the evolution of LUT parameters and the Jarlskog invariant.   Interestingly, the geometric  approach based on LUT allows us to express the oscillation probabilities for a given pair of neutrino flavours  in terms of only three (and not four) degrees of freedom which are related to the geometric properties  (sides and angles) of the triangle. Moreover, the LUT parameters are invariant under rephasing transformations and independent of the parameterization  adopted. }}}

\preprint{CTPU-PTC-21-01}
\maketitle

\section{Introduction}

Most of the  experimental neutrino oscillation data, coming from neutrinos produced at natural and artificial sources such as the Sun, Earth's atmosphere, nuclear reactors  or particle accelerators, can be   accommodated within the three-neutrino framework~\cite{deSalas:2020pgw, Esteban:2020cvm, Capozzi:2017ipn}. 
Along their path towards detectors, neutrinos can propagate through vacuum or through matter. For propagation in vacuum, the Schr{\"o}dinger equation describing neutrino oscillations can be solved exactly.
However, when neutrinos propagate through matter, they experience an effective potential and, as a result, neutrino flavour oscillations may get dramatically impacted  due to the Mikheyev-Smirnov-Wolfenstein (MSW) effect~\cite{Wolfenstein:1977ue,Mikheev:1986gs,Mikheev:1986wj}. 
In this case, it becomes challenging to solve the neutrino evolution equation, especially if the matter density of the medium traversed by neutrinos varies along the neutrino path, as it happens in the propagation through the Sun or the Earth.
If one assumes the density to be constant for a given baseline, analytic solutions can be found~\cite{Zaglauer:1988gz}. 
However, these expressions are quite complicated and do not allow a deeper understanding of the underlying structure of three-neutrino oscillations.
To facilitate this understanding, several approximate analytic expansions of the probability functions have been proposed~\cite{Arafune:1996bt,Cervera:2000kp,Freund:2001pn,Akhmedov:2004ny,Minakata:2009sr,Asano:2011nj,Li:2016txk,Minakata:2015gra,Denton:2016wmg,Denton:2018hal,Ioannisian:2018qwl} using different expansion parameters.
In a recent study~\cite{Parke:2019vbs}, the authors have explored the regime of validity  of these approximate expressions and presented a comparison among them.  

The above  holds for neutrino interactions allowed within the Standard Model, referred to as standard interactions (SI). Although the neutrino experiments are currently entering the precision era~\cite{deSalas:2020pgw}, the available data still cannot exclude the presence of (possibly large) non-standard neutrino interactions (NSI). For recent reviews on NSI see, for example, Refs.~\cite{Farzan:2017xzy, Esteban:2018ppq, Dev:2019anc}. 
Since the size of neutrino NSI with matter is proportional to the medium density,
obtaining  analytical expressions of oscillation probabilities in presence of NSI is a cumbersome exercise. Work in this direction can be found in Refs.~\cite{Friedland:2006pi,
Kikuchi:2008vq,Meloni:2009ia,
Asano:2011nj,Liao:2016hsa,Chaves:2018sih}.

In the usual description of three-flavour neutrino oscillations (including the case of NSI), there are four {\hlpm{independent}} degrees of freedom. In the commonly adopted PMNS parameterization, these are the three angles $\theta_{12}$, $\theta_{23}$ and $\theta_{13}$ which have been measured fairly well~\cite{deSalas:2020pgw} and the Dirac CP phase $\delta_{}$ which is related to the amount of CP violation in the leptonic sector. However, there is a strong dependence on the parameterization used which could lead to serious implications on our inferences about CP violation in the leptonic sector~\cite{Denton:2020igp}.  {\hlpm{It should be noted that these parameters could also be four phases~\cite{Aleksan:1994if} and need not be three real angles and one phase }}.

Recently, there is renewed interest in visualizing neutrino oscillations, which yields an alternative geometrical viewpoint towards our understanding of CP violation effects. While the two flavor case can  be conveniently visualized using the two-dimensional Poincare (or Bloch) sphere~\cite{Mehta:2009ea}, the generalization to a $N (\ge 3)$ level system is not straightforward~\cite{KIMURA2003339}.  Unitarity triangles are one of the widely studied approaches both in the quark sector as well as in the leptonic sector. It is important to note that the parameters of the unitarity triangle are invariant under rephasing transformations~\cite{Jarlskog:1985ht,PhysRevLett.55.2935,Jarlskog:2004be,Nieves:1987pp,Jenkins:2007ip} and independent of the parametrization adopted~\cite{Denton:2020igp}.
The next generation of neutrino oscillation experiments are expected to measure some of the elements of the mixing matrix to ${\cal O}$(1\%) precision. This would allow for precision tests of  unitarity. Likewise, the construction of a leptonic unitarity triangle (LUT) may be well within our reach.

Some interesting aspects related to LUT have been studied in  the literature. {\hlpm{The idea of unitarity triangles in the leptonic sector was first discussed in~\cite{Fritzsch:1999ee}.   Conditions for the observability of CP violation in terms of the LUT were discussed in~\cite{Sato:2000wv}.}}
The authors of~\cite{AguilarSaavedra:2000vr}  studied CP violation in the leptonic sector with Majorana neutrinos and interpreted its well-known features in terms of geometrical properties of unitarity polygons  for the three-flavour case and beyond.
The possibility of reconstructing the unitarity triangle in future oscillation and non-oscillation experiments was considered in Ref.~\cite{Farzan:2002ct}, where a set of measurements potentially allowing  to measure all sides of the triangle, and consequently to establish CP violation, was suggested. 
In~\cite{Zhang:2004hf}, the authors presented a study of LUT   and obtained analytical expressions for the sides and inner angles of the unitary triangle in vacuum, as well as its counterpart in matter. 
A simplified version of the unitarity triangles, more convenient to establish direct comparison with the experiments, was obtained in Ref.~\cite{Bjorken:2005rm} for the particular case of trimaximal mixing for the $\nu_2$ mass eigenstate.
The same choice for the neutrino mixing in connection with the LUT was explored in \cite{Ahuja:2007cu}. {\hlpm{The connection between a specific neutrino mass matrix texture and parameters of the Majorana unitarity triangle was explored in~\cite{Verma:2018yiu}.}}
He and Xu expressed the neutrino oscillation probabilities in terms of the parameters of the LUT in vacuum~\cite{He:2013rba} as well as in matter, assuming standard interactions~\cite{He:2016dco}. In a more recent study~\cite{Ellis:2020ehi}, a comprehensive analysis of LUT was presented, using both current neutrino oscillation data and projections of next-generation oscillation measurements. {\hlpm{For a recent review, see~\cite{Xing:2019vks}.}}
Finally, in Ref.~\cite{Dueck:2010fa},  the idea of  unitarity boomerangs  put forth in the quark sector~\cite{FRAMPTON201067}  was extended  to the leptonic sector.

The idea of the present work is to go beyond the study carried out in the context of vacuum~\cite{He:2013rba}  and matter (assuming standard interactions)~\cite{He:2016dco} and obtain analytical expressions for probabilities in the $\nu_\mu\to\nu_e$ and $\nu_\mu \to \nu_\tau$ channels~\footnote{$\nu_\tau$ appearance has been discussed in the context of the upcoming DUNE experiment in~\cite{deGouvea:2019ozk,Ghoshal:2019pab,Machado:2020yxl,Rout:2020cxi}.} in presence of NSI. Then, we will test the accuracy of our expression and construct the LUT parameters for specific channels. We will study the evolution of LUT as a function of the neutrino energy for a fixed baseline of $1300$ km. We will also analyze the dependence of the Jarlskog invariant on the nature of interactions.  To the best of our knowledge,  a study of neutrino oscillations in presence of a new physics scenario such as NSI  using the geometric approach based on LUT has not been  reported  earlier.

The paper is structured as follows: in Sec.~\ref{sec:analysis}, we develop the idea of the geometric interpretation of neutrino oscillations in vacuum, standard matter and NSI. We derive an approximate formula for neutrino oscillations in presence of NSI and show that the oscillation probability can be expressed in terms of the parameters of the LUT.  Next, we test the validity of the expressions obtained in this work and those existing in literature in Sec.~\ref{sec:validity}. In Sec.~\ref{sec:evolve}, we study numerically  the evolution of the parameters of the LUT  as a function of the energy in  presence of NSI. In Sec.~\ref{sec:jcp}, we  study  the evolution of the Jarlskog invariant, $J_{CP}$, for the NSI case. Finally, we conclude in Sec.~\ref{sec:conclude}.

\section{Geometrical interpretation of neutrino oscillations}
\label{sec:analysis}

%
Three-generation neutrino mixing can be described by a $3\times 3 $ unitary mixing matrix $U$, which appears in the weak charged current interactions. The commonly used form given below is referred to as the  Pontecorvo-Maki-Nakagawa-Sakata (PMNS) parametrization~\cite{Zyla:2020zbs}
\bea 
U \equiv 
{U}_{\text{PMNS}} 
&=& \left(
\begin{array}{ccc}
1   & 0 & 0 \\  0 & c_{23}  & s_{23}   \\ 
 0 & -s_{23} & c_{23} \\
\end{array} 
\right)   
  \left(
\begin{array}{ccc}
c_{13}  &  0 &  s_{13} e^{- i \delta}\\ 0 & 1   &  0 \\ 
-s_{13} e^{i \delta} & 0 & c_{13} \\
\end{array} 
\right)  \left(
\begin{array}{ccc}
c_{12}  & s_{12} & 0 \\ 
-s_{12} & c_{12} &  0 \\ 0 &  0 & 1  \\ 
\end{array} 
\right)  \,,
\label{eq:u}
 \eea 
where $s_{ij}=\sin {\theta_{ij}}, c_{ij}=\cos \theta_{ij}$ and $\delta$ is the Dirac-type CP phase. If neutrinos are Majorana particles, there can be two additional Majorana-type phases in the three flavour case, as $U \to U$ diag$(1, e^{i \kappa}, e^{i \zeta} ) $. However, these Majorana phases play no role in neutrino oscillation studies as they give rise to an overall phase in the neutrino oscillation amplitude which is not measurable. 
The unitarity of the mixing matrix, $U^{\dagger}U = UU^{\dagger} = {\mathbb{I}}_{3\times 3}$, leads to  the so-called leptonic unitarity triangles in the complex plane, 
\bea 
\label{diraclut}
\sum_j U_{\ell j} U^\star _{\ell' j} = 
U_{\ell1}^{} U^{*}_{\ell'1} + U_{\ell2} ^{} U^{*}_{\ell'2} + U_{\ell3}^{} U^{*}_{\ell'3} &=& 0 \quad {\rm{with}} \quad l \neq l^\prime\,.
\eea
\bea
\sum_l U_{\ell j} U^\star _{\ell j'} = 
U_{e j}^{} U^{*}_{ e j'} + U_{\mu j}^{} U^{*}_{\mu j'} + U_{\tau j}^{} U^{*}_{\tau j'} &=& 0 \quad {\rm{with}} \quad j \neq j^\prime\,. 
\label{eq:1}
\eea
Here, Eq.~(\ref{diraclut}) represents row or Dirac triangles while Eq.~(\ref{eq:1}) represents  column or Majorana triangles. 
%
\begin{figure}[t!]
\centering
\includegraphics[scale=0.8]{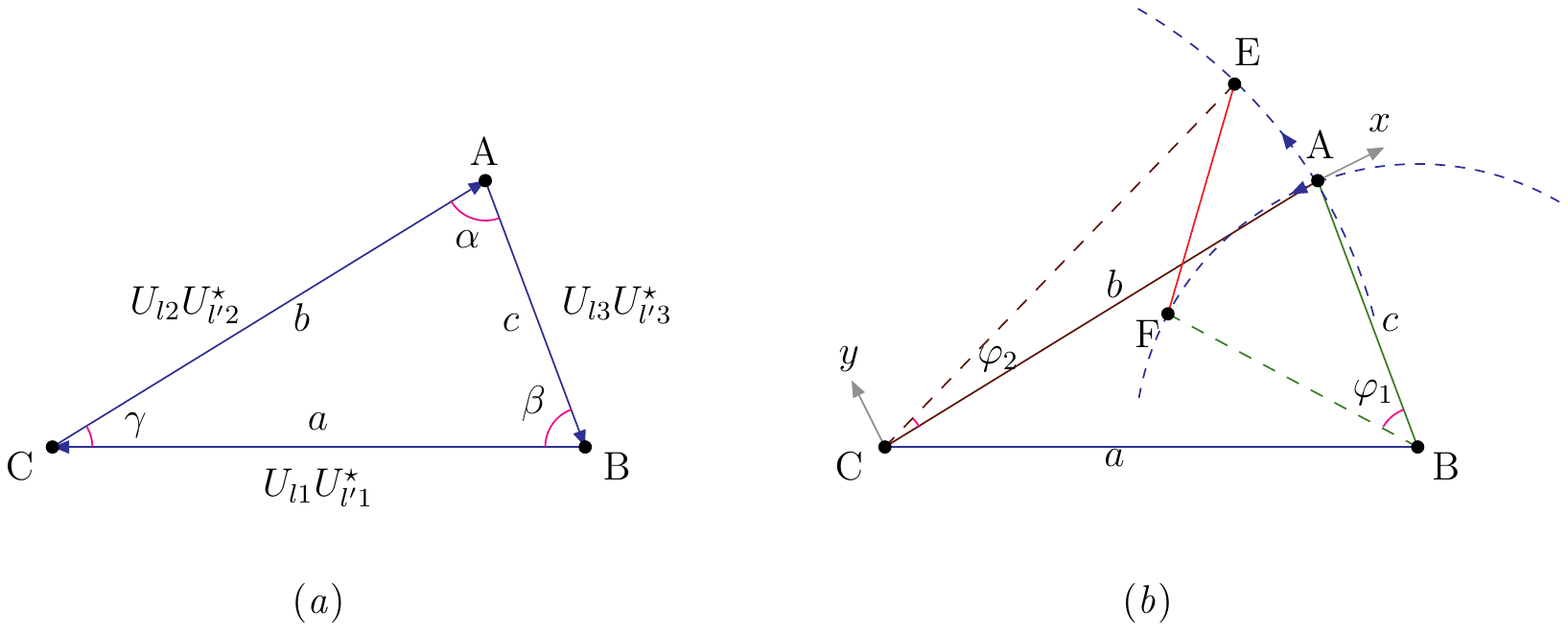}
\vskip -6in
\caption{
(a) Unitarity triangle with the sides $(a,b,c)$ and angles $(\alpha,\beta,\gamma)$. 
(b) Modification of the unitarity triangle due to neutrino propagation
(adapted from~\cite{He:2016dco}).
}
\label{fig:uni_tri}
\end{figure}
Since we are interested in neutrino flavour oscillations, we will consider only the Dirac LUT, as shown in Fig.~\ref{fig:uni_tri}(a). The sides $(a,b,c)$ and angles $(\alpha,\beta,\gamma)$ of the LUT can be directly expressed in terms of the elements of the leptonic mixing matrix, $U$~\cite{He:2016dco}
\begin{align}
\left(a, b, c \right) &= \left(|U_{\ell1}^{} U^{*}_{\ell'1}|, |U_{\ell2}^{} U^{*}_{\ell'2}|, |U_{\ell3}^{} U^{*}_{\ell'3}|\right) ~, \nonumber \\
\left(\alpha, \beta, \gamma\right) &= \left(\arg \left[-\frac{U_{\ell3} ^{} U^{*}_{\ell'3}}{U_{\ell2}^{} U^{*}_{\ell'2}}\right],
\arg \left[-\frac{U_{\ell1}U^{*}_{\ell'1}}{U_{\ell3}U^{*}_{\ell'3}}\right], 
\arg\left[-\frac{U_{\ell2}U^{*}_{\ell'2}}{U_{\ell1}U^{*}_{\ell'1}}\right] \right)~.
\label{eq:lut_defn}
\end{align}
As stated in~\cite{AguilarSaavedra:2000vr} (see also~\cite{Nieves:1987pp,Jenkins:2007ip}), under the allowed rephasing transformations, $l_{Lj,Rj} \to e^{i \lambda_j}l_{Lj,Rj}$, the mixing matrix elements transform as 
$U_{lj} ^{} \to e^{i\lambda_j} U_{lj}^{}$ and, thus, the minimal rephasing invariant terms are the products ($U_{lj}^{} U_{l'j}^\star$) with the minimal CP violating quantities Im($U_{lj}^{} U_{l'j}^\star$), subject to the requirement that they need to interfere with their real part or with other sides of the triangle.  

Note that, out of the six degrees of freedom appearing in Eq.~\eqref{eq:lut_defn}, only three taken to be two sides and one angle enter the probability expression for a given appearance channel, i.e. with a given pair ($l,l'$), as we will see below. 
For a different appearance channel, the parameters will no longer remain the same, as expected. It is interesting to note the connection between LUT parameters and the Jarlskog factor~\cite{Jarlskog:1985ht,PhysRevLett.55.2935,Jarlskog:2004be}, $J_{CP} = {\textrm{Im}}\left(U_{\ell' j}^{} U_{\ell j}^\star U_{\ell j'} ^\star U_{\ell' j'} ^ {} \right)$ where $ l \neq l^\prime$ and $j \neq j^\prime$~\cite{He:2013rba}. 
\begin{align}
J_{CP} 
  &= bc\sin\alpha = ca \sin \beta = ab \sin \gamma~.
\label{eq:lut_j}
\end{align} 
As a consequence of the orthogonality of any pair of different rows or columns of the mixing matrix, it turns out that {the absolute value of} $J_{CP}$ is unique { for all of them,} and can  differ at most by a sign.  {\hlpm{It is interesting to note that  the vacuum and  matter (with standard interactions) counterpart of $J_{CP}$ are related~\cite{Naumov:1991ju,HARRISON2000349,Wang:2019yfp,Denton:2019yiw}.}}

\subsection{Neutrino propagation in vacuum}
For neutrino oscillations in vacuum, the Hamiltonian is given by
\be
H = \frac{\ldm}{2E}U\left(\begin{array}{ccc} 0 & 0 & 0 \\ 0 & r & 0 \\ 0 & 0 & 1 \\\end{array}\right)U^{\dagger}.
\ee
where $r = {\sdm}/{\ldm}$.
The oscillation probability in terms of {three} independent LUT parameters ($b$, $c$ and $\alpha$), defined in Eq.~\eqref{eq:lut_defn}, is given by~\cite{He:2016dco}
\be
P_{\ell\ell'} = 4c^{2}\sin^{2}\Delta_{31} - 8bc\sin\Delta_{31} \sin \Delta_{21} \cos[\Delta_{32} + \alpha] + 4b^{2}\sin^{2}\Delta_{21}~,
\label{eq:osc_vac}
\ee
where $\Delta_{jk} = {\Delta m^{2}_{jk} L}/{4E}$.
As a consequence of neutrino propagation, the vector $\protect\overrightarrow{CA}$ of the LUT rotates by an angle $\varphi_{2} = 2\Delta_{21}$ and the vector $\protect\overrightarrow{BA}$ rotates by $\varphi_{1} = 2\Delta_{31}$ giving rise to a quadrilateral CEFB, see Fig.~\ref{fig:uni_tri}(b). It can also be shown that $P_{\ell\ell'} = |\overrightarrow{EF}|^{2}$~\cite{He:2016dco}.

\subsection{Neutrino propagation in matter  with standard interactions}

We next consider the Hamiltonian describing neutrino evolution in constant matter governed by SI, given by
\bea
\label{eq:h_si}
H &=& \frac{\ldm}{2E}U\left(\begin{array}{ccc} 0 & 0 & 0 \\ 0 & r & 0 \\ 0 & 0 & 1 \\\end{array}\right)U^{\dagger} + \sqrt{2}G_{F}N_{e}\left(\begin{array}{ccc} 1 &0 & 0 \\ 0 &0& 0 \\ 0 & 0 &0\\ \end{array}\right) = \frac{\Delta m_{31}^{2}}{2E} \left[U K U^{\dagger}\right]~,
\eea
where the dimensionless matrix $K$ takes the form
\bea\label{eq:k_si}
 K&=&\left(\begin{array}{ccc}0&0 & 0\\ 0 &r&0\\0 &0&1\end{array} \right) 
+ U^{\dagger}\left(\begin{array}{ccc}A&0 & 0\\ 0 &0&0\\0 &0&0\end{array} \right)U~.
\eea
Here, $A$ is given by 
\bea 
\label{eq:def_nE}
A &=& \dfrac{2\sqrt{2}G_{F}N_{e}E}{\Delta m^{2}_{31}} {\textcolor{black}{ = 2 \times 0.76 \times 10^{-4} \times Y_e {\dfrac{\rho}{[\text{g/cc}]}  \dfrac{E}{[\text{GeV}]}}\dfrac{[\text{eV}^2]}{\Delta m^{2}_{31}}}}~.
\eea
{\textcolor{black}{where, $Y_e$ is the electron fraction ($Y_e \simeq 0.5$ for earth matter) and $\rho$ is the earth matter density.} As it was shown in Ref.~\cite{Zaglauer:1988gz}, the modified neutrino oscillation parameters in matter can be calculated exactly for constant matter density. 
The neutrino oscillation probability in matter can be written down in terms of parameters of LUT as~\cite{He:2016dco}
\be
P_{\ell\ell'} = 4(c^{m})^{2}\sin^{2}\Delta_{31}^m - 8b^{m}c^{m}\sin\Delta_{31}^m \sin \Delta_{21}^m \cos[\Delta_{32}^m + \alpha^{m}] + 4(b^{m})^{2}\sin^{2}\Delta_{21}^m~,
\label{eq:prob_si}
\ee
where  the mixing matrix in vacuum is replaced by the mixing matrix in matter, $U \to U^m$. Therefore,   the three independent LUT parameters   get modified, $b \to b^m$, $c\to c^m$ and $\alpha \to \alpha^m$. Note that the form of Eq.~\eqref{eq:prob_si} is similar to Eq.~\eqref{eq:osc_vac}.
The mixing angles, CP phase and mass-squared splittings in matter can be found in Ref.~\cite{Zaglauer:1988gz}\footnote{Note, however, that there are two typos in the original reference, see for example Sec.~4.2.1 of Ref.~\cite{Parke:2019vbs} for a corrected version.}.
%

\subsection{Neutrino propagation in matter with non-standard interactions}
\label{sec:nsi}

Neutrino NSI can be parameterized as 4-fermion terms in the Lagrangian density. They can induce new charged current (CC) and  neutral current (NC) interactions, given by

\bea
\mathcal{L}_{\text{CC}} &=& -2\sqrt{2}G_F\varepsilon_{\alpha\beta}^{ff'X}(\overline{\nu}_\alpha\gamma^\mu P_L l_\beta)(\overline{f}'\gamma_\mu P_X f)~,
\\
\mathcal{L}_{\text{NC}} &=& -2\sqrt{2}G_F\varepsilon_{\alpha\beta}^{fX}(\overline{\nu}_\alpha\gamma^\mu P_L \nu_\beta)(\overline{f}\gamma_\mu P_X f)~.
\eea
The dimensionless coefficients $\varepsilon_{\alpha\beta}^{f(f')X}$ quantify the strength of the NSI with respect to the standard weak interaction.
Here $f$ and $f'$ refer to the charged fermions involved in the interactions (electrons and up and down-quarks), while $X$ denotes the left and right chirality of the projection operator $P_X$.
The CC NSI can affect the neutrino production and detection processes. 
In this work, however, we focus only on NC NSI, since we are interested in the effect of NSI in the neutrino propagation through matter, which is affected by the vector part of the NSI, given by $\varepsilon_{\alpha\beta}^{fV} = \varepsilon_{\alpha\beta}^{fL} + \varepsilon_{\alpha\beta}^{fR} $.
In the most general case, the neutrino oscillation probability is indeed affected by the combination
\begin{equation}
 \varepsilon_{\alpha\beta} =  \varepsilon_{\alpha\beta}^{eV} + \frac{N_u}{N_e}\varepsilon_{\alpha\beta}^{uV} + \frac{N_d}{N_e}\varepsilon_{\alpha\beta}^{dV}\,,
\end{equation}
with the electron, up-quark and down-quark densities denoted by $N_e$, $N_u$ and $N_d$, respectively. 
The Hamiltonian with NSI is then given by,
\bea\label{eq:h_nsi}
H &=& \frac{\ldm}{2E}U\left(\begin{array}{ccc} 0 & 0 & 0 \\ 0 & r & 0 \\ 0 & 0 & 1 \\\end{array}\right)U^{\dagger} + \sqrt{2}G_{F}N_{e}\left(\begin{array}{ccc} 1 &0 & 0 \\ 0 &0& 0 \\ 0 & 0 &0\\ \end{array}\right) 
+ \sqrt{2}G_{F}N_{e} {\underbrace{\left(\begin{array}{ccc} \eee & \eem & \eet \\
\eem^{*} & \emm & \emt \\
\eet^{*} & \emt^{*} 
& \ett
 \end{array} \right)}_{\varepsilon}} \nonumber \\
&=& \frac{\ldm}{2E}U^{m}
\bigg[
K_{\text{diag}} + A \tilde{\varepsilon}
\bigg](U^{m})^{\dagger}  ~,
\eea
where $K_{\text{diag}}$ contains the eigenvalues for the SI case and $\tilde{\varepsilon} = (U^{m})^{\dagger}\varepsilon U^{m}$. 
In analogy with the matrix $K$ in Eq.~\eqref{eq:k_si}, here we define the corresponding matrix in the presence of NSI as 
\be\label{eq:k_nsi}
K^{\text{NSI}} = K_{\text{diag}} + A \tilde{\varepsilon} ~.
\ee 
In order to find the effective mass-squared splittings in the case of NSI, we can diagonalise $K^{\text{NSI}}$ with a matrix $S$ such that $S^{\dagger}K^{\text{NSI}}S = K^{\text{NSI}}_{\text{diag}}$ and obtain a modified mixing matrix   given by $U^{N} = U^{m}S$. 
We then introduce NSI as a perturbation to the standard matter potential and invoke time-independent 
perturbation theory to diagonalise the perturbed matrix $K^{\text{NSI}}$.
For a perturbed matrix, $H = H^{0} + \lambda H^{(1)}$, the modified eigenvalues are given (after first order correction) by 
$E_{n} = E_{n}^{(0)} + \lambda \bra{\Phi_{n}^{(0)}} H^{(1)} \ket{\Phi_{n}^{(0)}}$,  
where $E_{n}^{(0)}$ is the unperturbed eigenvalue with the corresponding unperturbed eigenvector being  $\ket{\Phi_{n}^{(0)}}$.
The modified eigenvectors of $H$, after first order correction, are given by, 
\be
\ket{\Phi_{n}} = \ket{\Phi_{n}^{(0)}} - \lambda \sum_{k \neq n} 
\frac{\bra{\Phi_{k}^{(0)}} H^{(1)} \ket{\Phi_{n}^{(0)}}}{E_{k}^{(0)} - E_{n}^{(0)}} 
\ket{\Phi_{k}^{(0)}}~.
\ee

Using these, the relevant elements of $U^{N}$   can be expressed in terms of  elements of $U^m$ as
\bea\label{eq:u_nsi_pert}
U_{12}^{N} &=& N_{2}\big[U_{12}^{m} 
- A\frac{\tilde{\varepsilon}^{*}_{23}}{\lambda_{3}-\lambda_{2}}U_{13}^{m}
- A\frac{\tilde{\varepsilon}_{12}}{-\lambda_{2}}U_{11}^{m}\big]~, ~ 
U_{13}^{N} = N_{3}\big[U_{13}^{m} 
- A\frac{\tilde{\varepsilon}_{23}}{\lambda_{2}-\lambda_{3}}U_{12}^{m}
- A\frac{\tilde{\varepsilon}_{13}}{-\lambda_{3}}U_{11}^{m}\big]~, \nonumber \\
U_{22}^{N} &=& N_{2}\big[U_{22}^{m} 
- A\frac{\tilde{\varepsilon}^{*}_{23}}{\lambda_{3}-\lambda_{2}}U_{23}^{m}
- A\frac{\tilde{\varepsilon}_{12}}{-\lambda_{2}}U_{21}^{m}\big]~, ~
U_{23}^{N} = N_{3}\big[U_{23}^{m} 
- A\frac{\tilde{\varepsilon}_{23}}{\lambda_{2}-\lambda_{3}}U_{22}^{m}
- A\frac{\tilde{\varepsilon}_{13}}{-\lambda_{3}}U_{21}^{m}\big]~, \nonumber \\
U_{32}^{N} &=& N_{2}\big[U_{32}^{m} 
- A\frac{\tilde{\varepsilon}^{*}_{23}}{\lambda_{3}-\lambda_{2}}U_{33}^{m}
- A\frac{\tilde{\varepsilon}_{12}}{-\lambda_{2}}U_{31}^{m}\big]~, ~
U_{33}^{N} = N_{3}\big[U_{33}^{m} 
- A\frac{\tilde{\varepsilon}_{23}}{\lambda_{2}-\lambda_{3}}U_{32}^{m}
- A\frac{\tilde{\varepsilon}_{13}}{-\lambda_{3}}U_{31}^{m}\big]~, \nonumber \\
\eea
where $\lambda_j = (\Delta m^2_{j1})^m / \Delta m^2_{31}$. $N_2$ and $N_3$ are normalisation factors, given by
\bea\label{eq:pert_evec_norm}
N_{2} &=& \bigg[1 + \frac{A^{2}|\tilde{\varepsilon}_{23}|^{2}}{(\lambda_{3}-\lambda_{2})^{2}} 
+ \frac{A^{2}|\tilde{\varepsilon}_{12}|^{2}}{\lambda_{2}^{2}}\bigg]^{-1/2}, \nonumber \\
N_{3} &=& \bigg[1 + \frac{A^{2}|\tilde{\varepsilon}_{23}|^{2}}{(\lambda_{3}-\lambda_{2})^{2}} 
+ \frac{A^{2}|\tilde{\varepsilon}_{13}|^{2}}{\lambda_{3}^{2}}\bigg]^{-1/2}.
\eea
Once we have computed the elements of $U^N$, we can readily find the NSI modified LUT parameters using Eq.~\eqref{eq:lut_defn} with $U\to U^N$,
\begin{align}
\label{eq:lut_defn_nsi}
&b^{N}_{\mu e} = |U^N_{22}U^{N*}_{12}|~,\quad 
c^{N}_{\mu e} = |U^N_{23}U^{N*}_{13}|~,\quad
\alpha^{N}_{\mu e} = \arg\left(-\frac{U^N_{23}U^{N*}_{13}}{U^N_{22}U^{N*}_{12}}\right)~;
\nonumber \\
&b^{N}_{\mu \tau} = |U^N_{22}U^{N*}_{32}|~,\quad 
c^{N}_{\mu \tau} = |U^N_{23}U^{N*}_{33}|~,\quad
\alpha^{N}_{\mu \tau} = \arg\left(-\frac{U^N_{23}U^{N*}_{33}}{U^N_{22}U^{N*}_{32}}\right)~.
\end{align}
Note that the subscripts indicate the oscillation channel ($\nu_{\mu} \to \nu_{e}$ or $\nu_{\mu} \to \nu_{\tau}$) under consideration. {{It is worth pointing out that $b^N_{\alpha \beta}, c^N_{\alpha \beta}, \alpha^N_{\alpha\beta}$ depend only on the elements of the mixing matrix in a rephasing invariant form~\cite{AguilarSaavedra:2000vr} and are independent of  the specific parameterization.}}
The {first order} perturbed eigenvalues 
 of $K^{\text{NSI}}$ are given by,
\bea\label{eq:mass_nsi_pert}
(\ldm)^{N} &\simeq& (\ldm)^{m} + A(\epst{3}{3}-\epst{1}{1}) 
\ldm~, \nonumber \\
(\sdm)^{N} &\simeq&  (\sdm)^{m} + A(\epst{2}{2}-\epst{1}{1}) 
\ldm~.
\eea
The oscillation probability takes the same form as  in the case of standard matter (Eq.~\eqref{eq:prob_si}), or in the case of vacuum (Eq.~\eqref{eq:osc_vac}),  with the appropriately modified values of the sides and angles of the LUT relevant for the particular channel: 
\be
P_{\ell\ell'} = 4(c^{N}_{\ell\ell'})^{2}\sin^{2}\Delta_{31}^N 
- 8b^{N}_{\ell\ell'}c^{N}_{\ell\ell'}\sin\Delta_{31}^N \sin \Delta_{21}^N \cos[\Delta_{32}^N + \alpha^{N}_{\ell\ell'}] + 4(b^{N}_{\ell\ell'})^{2}\sin^{2}\Delta_{21}^N~.
\label{eq:prob_nsi}
\ee
Expressing the NSI oscillation probability as Eq.~\eqref{eq:prob_nsi} and thereby establishing the robustness of the expression of oscillation probability in terms of the LUT parameters  in the presence of NSI constitutes the key result of this work. Thus, the invariance of the probability expression allows for a neat geometric view of neutrino oscillations in terms of LUT, as depicted in Fig.~\ref{fig:uni_tri}.  

It should be noted that the above description  
only applies to  appearance channels ($l\neq l^\prime$). 
For the disappearance channels ($l=l^\prime$), there is no LUT and, therefore, we can infer the disappearance probabilities only indirectly by imposing the unitarity condition.
For example, for the muon disappearance channel, we can write $P_{\mu\mu} = 1 - P_{\mu e} - P_{\mu\tau}$.
\section{Validity of the approximate  LUT expression for NSI}
\label{sec:validity}
\begin{figure}
\centering
\includegraphics[width=\textwidth]{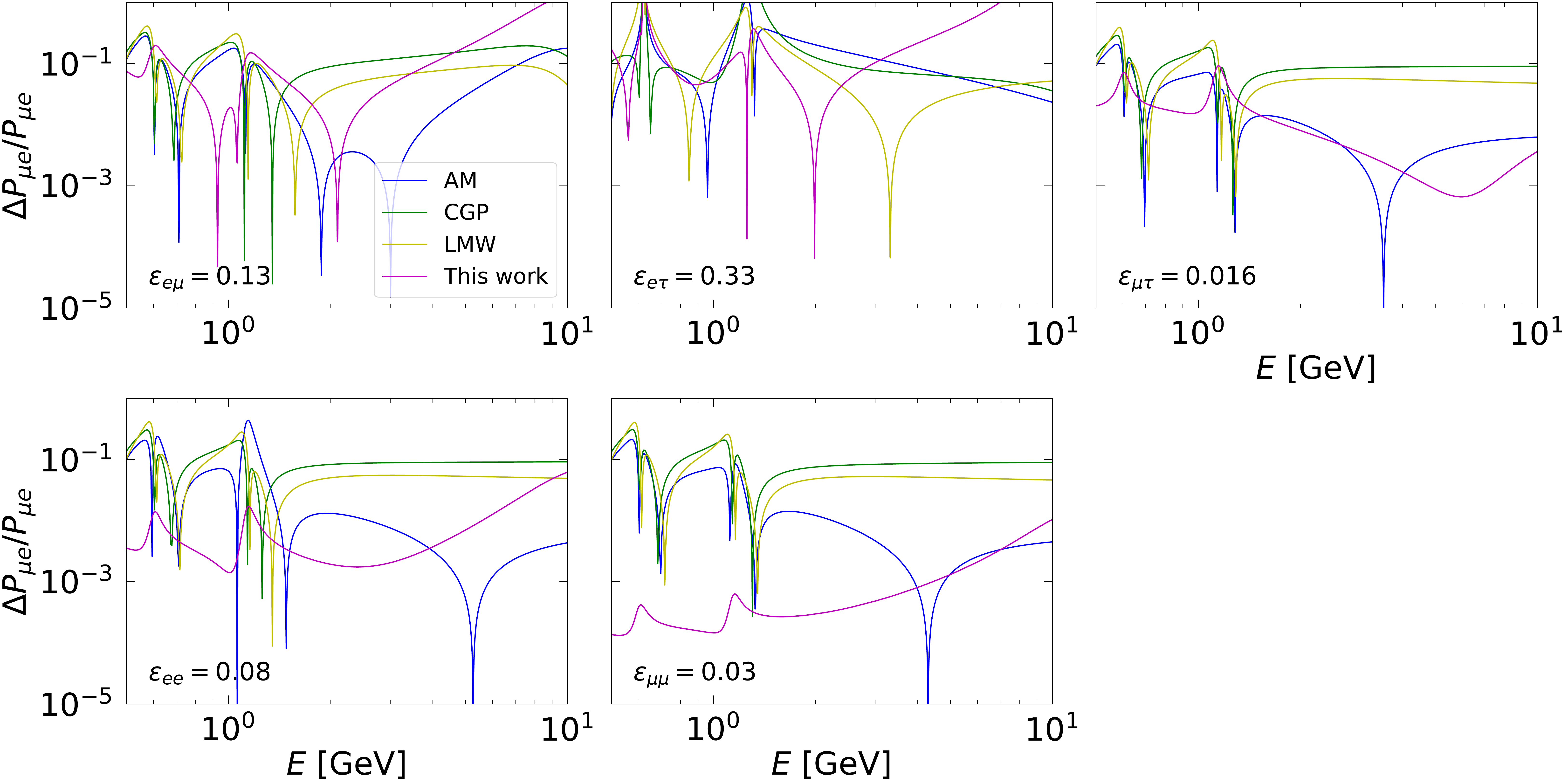}
\caption{Comparison of our approach with other formulas obtained in the literature. The fractional difference of probabilities ($\Delta P_{\mu e}/P_{\mu e}$) for different {{fixed values}} of $\varepsilon_{\alpha\beta}$  is plotted as a function of $E$. Each panel compares the precision for a single $\varepsilon_{\alpha\beta}$. In the case of the non-diagonal parameters, the phases are set to zero, $\varphi_{\alpha\beta} = 0$. The average value of density is taken to be $2.95$ g/cc which corresponds to $L \simeq 1300$ km.}
\label{fig:prob_precision} 
\end{figure}
\begin{figure}[htb]
\centering
\includegraphics[scale=0.6]{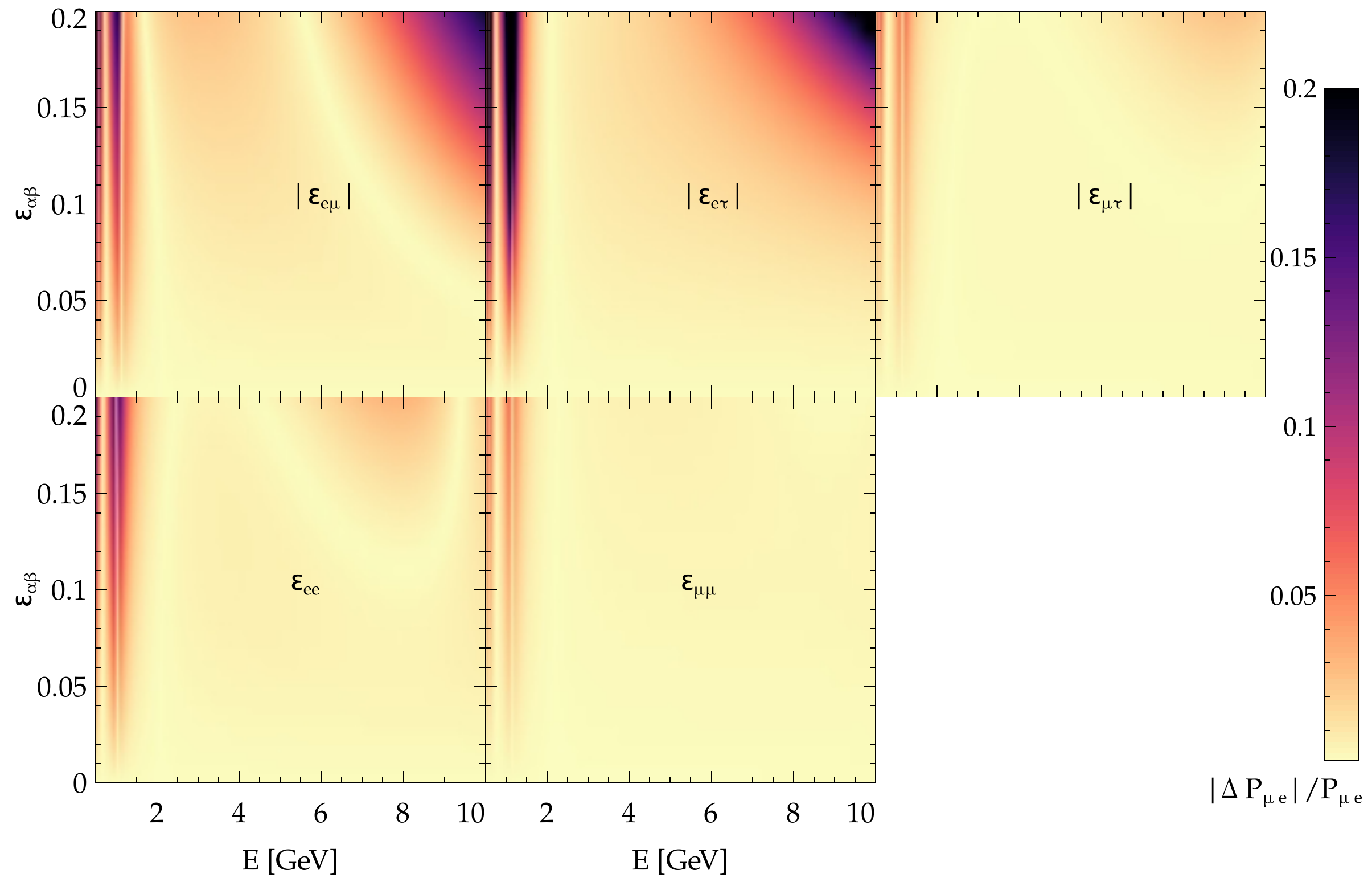}
\caption{ Heatmap for $|\Delta P_{\mu e}|/P_{\mu e}$   as a function of the NSI parameters $\eem, \eet, \emt, \eee, \emt$ (taken one-at-a-time and assuming them to be real) and the neutrino energy, $E$.}
\label{fig:prob_heatmap_mue}
\end{figure}
\begin{figure}[htb]
\centering
\includegraphics[scale=0.6]{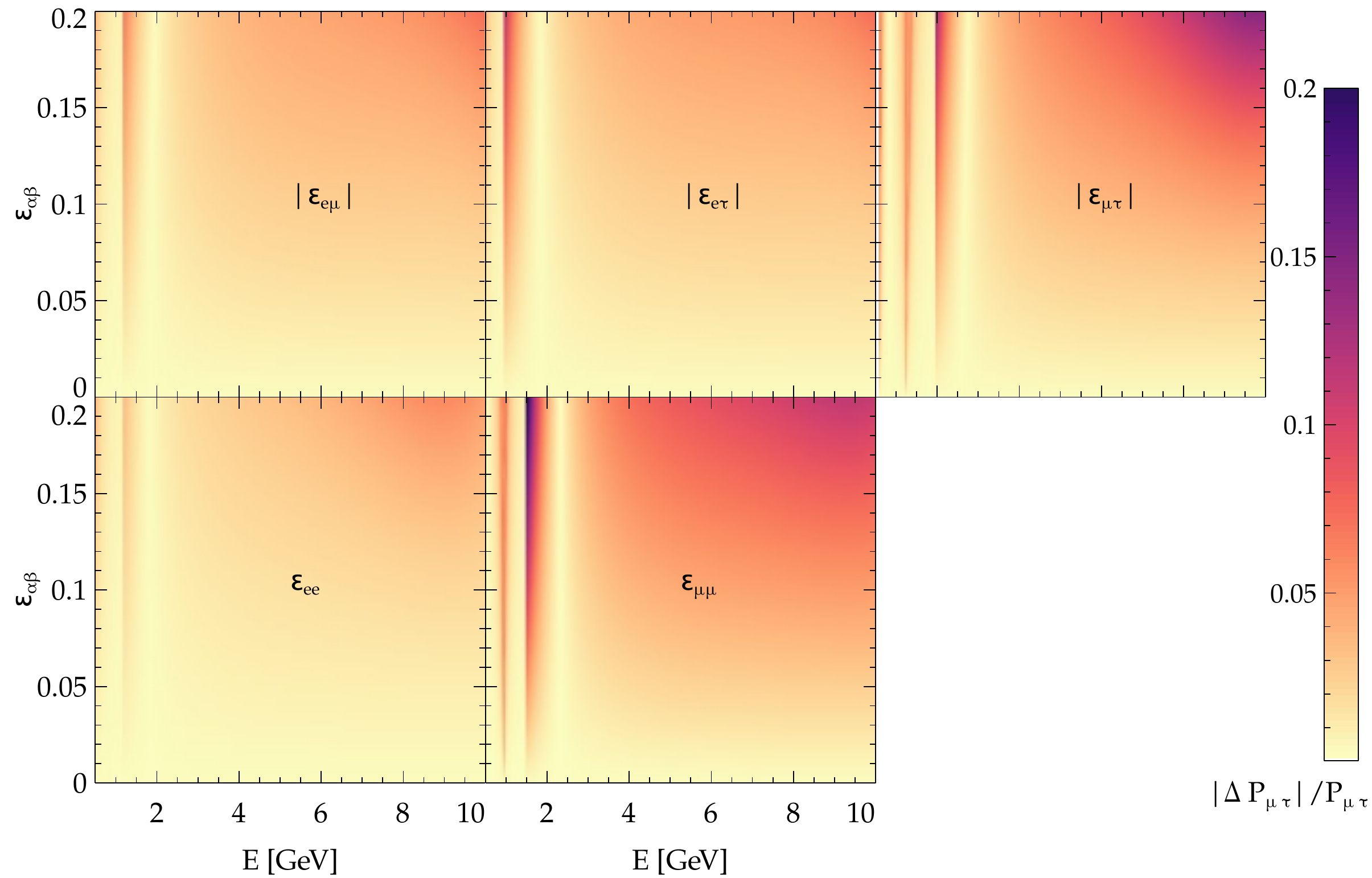}
\caption{Heatmap for $|\Delta P_{\mu \tau}|/P_{\mu \tau}$ as a function of the NSI parameters $\eem, \eet, \emt, \eee, \emt$ (taken one-at-a-time and assuming them to be real) and the neutrino energy, $E$. }
\label{fig:prob_heatmap_mutau}
\end{figure}
\begin{figure}[tb]
\centering
\includegraphics[scale=0.55]{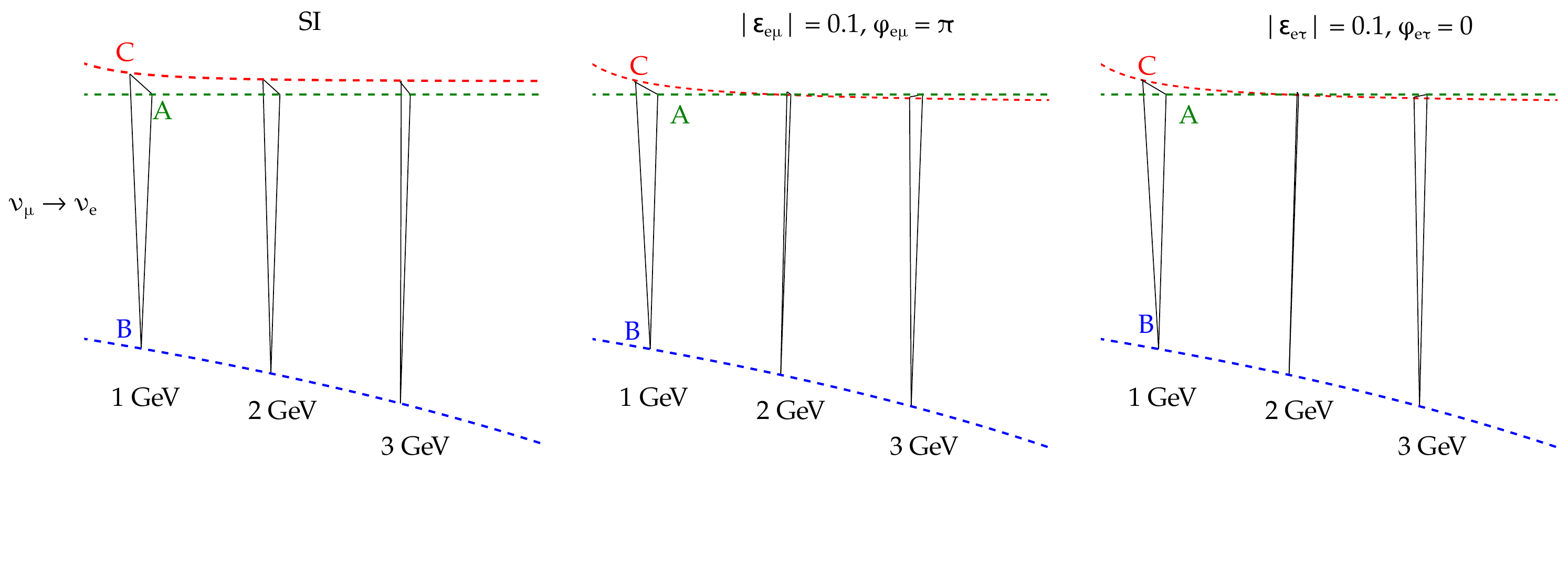}
\includegraphics[scale=0.55]{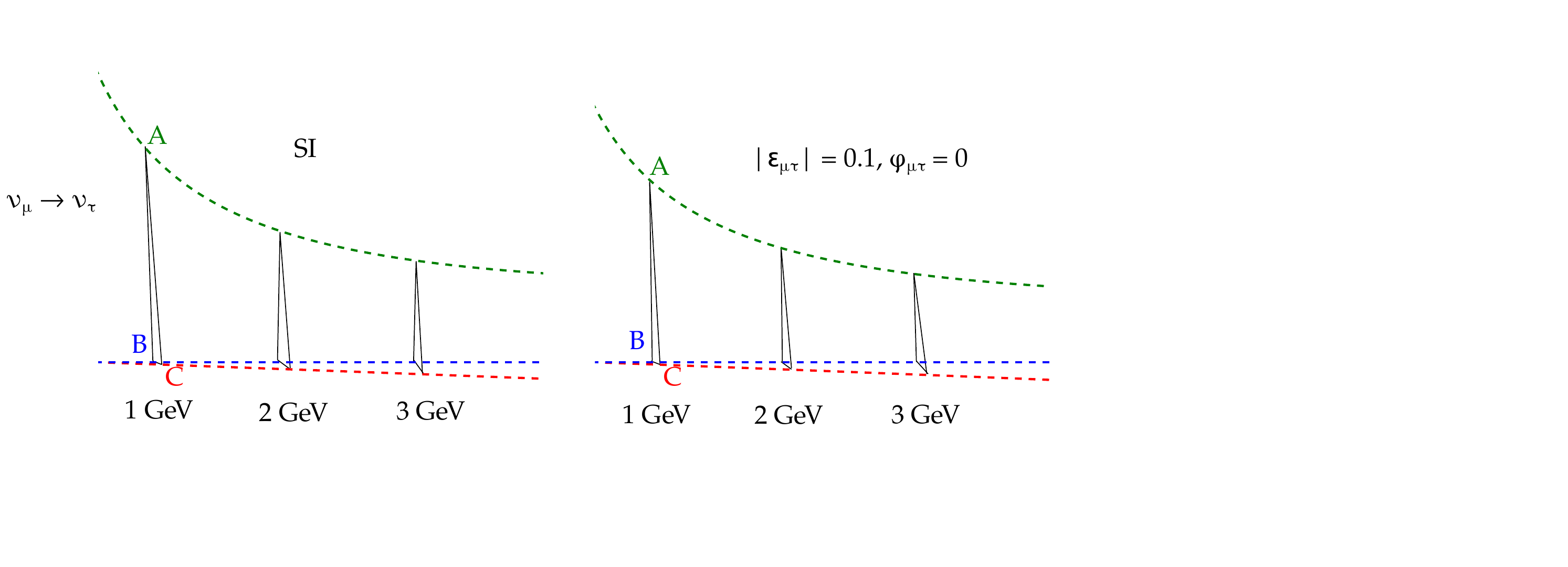}
\caption{Evolution of the LUT for the $\nu_{\mu} \to \nu_{e}$ (top row) and $\nu_{\mu} \to \nu_{\tau}$ channel (bottom row). 
The green, blue and red dashed lines  show the evolution of the three vertices $A$, $B$ and $C$  with the energy for the standard case (left) and the NSI case (middle and right panels).
}
\label{fig:evol}
\end{figure}

The accuracy of the LUT expression for the neutrino oscillation probability with standard matter interactions given by Eq.~\eqref{eq:prob_si} with respect to the exact numerical calculation was tested in Ref.~\cite{He:2016dco} for different baselines ranging between $295-1300$ km.
In this section, we compare the precision of our LUT formula in presence of NSI, Eq.~\eqref{eq:prob_nsi},  with existing expressions in the literature, taken from Ref.~\cite{Asano:2011nj} (AM)\,\footnote{We use Eq.~(36). of \cite{Asano:2011nj}.}, Ref.~\cite{Liao:2016hsa} (LMW) and Ref.~\cite{Chaves:2018sih} (CGP). 
We consider a fixed baseline of $L \simeq 1300$ km, relevant in the context of long-baseline neutrino experiments, such as DUNE~\cite{Acciarri:2015uup}.
The average value of {matter density in this case} is taken to be $2.95$ g/cc.
In order to compare the accuracy of the different approximate expressions, we calculate the deviations with respect to the exact value of the probability,  calculated numerically using GLoBES~\cite{Huber:2004ka, Huber:2007ji} with the extension {\it snu.c}~\cite{Kopp:2006wp,Kopp:2007ne}, $\Delta P_{\alpha\beta} = P_{\alpha\beta}^\mathrm{exact} - P_{\alpha\beta}^\mathrm{approx}$.
The relative precision for the $\nu_e$ appearance channel, $|\Delta P_{\mu e}|/P_{\mu e}$, for different values of $\varepsilon_{\alpha\beta}$ (taken one at a time) is plotted in Fig.~\ref{fig:prob_precision}, as a function of the neutrino energy, $E$. 
For the standard oscillation parameters, we choose the best-fit values from Ref.~\cite{deSalas:2020pgw}, while the NSI parameters indicated in each panel are set to their $90\%$ bounds summarized in Ref.~\cite{Farzan:2017xzy}. 
We find that, at low energies, our formula is comparable in precision with the other approximate expressions obtained in the literature in case of $\varepsilon_{e\mu}$ and $\varepsilon_{e\tau}$, while doing slightly better for the other $\varepsilon$'s. For larger energies, however,  the expressions become less precise due to break down of the perturbation theory.

In order to check the validity regime of Eq.~\eqref{eq:prob_nsi} in the plane $\varepsilon_{\alpha\beta}-E$, we show heatplots of the relative precision in the probability calculation for the $\nu_\mu \to \nu_e$  channel ($|\Delta P_{\mu e}|/P_{\mu e}$) in  Fig.~\ref{fig:prob_heatmap_mue} and $\nu_\mu \to \nu_\tau$ channel ($|\Delta P_{\mu \tau}|/P_{\mu \tau}$) in Fig.~\ref{fig:prob_heatmap_mutau}. 
The heatplots are calculated considering the presence of five (real) NSI parameters, taken one at a time, as indicated by the legends in each panel.
In presence of $\eem$ and $\eet$, it is easy to note that, as long as the NSI parameters are below $0.1$ and the neutrino energy is below $8$ GeV,  our estimated probability remains within roughly $(5-10)\%$ of the GLoBES result.
Beyond this region, the results are less accurate as the perturbation theory breaks down at higher energies and for larger NSI couplings. 
The effect of $\eee$ and $\emt$ throughout the energy range explored in our analysis gives probabilities consistent with GLoBES (the error being roughly within $(2-7)\%$). 
Only for $\emt \gtrsim 0.2$ and $E \gtrsim 9$ GeV, the analysis starts showing mild deviations from the numerical value.
In Fig.~\ref{fig:prob_heatmap_mutau}, the heatmap for $|\Delta P_{\mu \tau}|/P_{\mu \tau}$ in the plane of $\varepsilon_{\alpha\beta}-E$ also shows good agreement with the numerical results.
Only the presence of $\emt$ or $\eee$ with a strength larger than $0.15$ starts showing a deviation of the order of or larger than $(6-8)\%$ at higher energies.

\section{Evolution of the Leptonic Unitary Triangle with the energy}
\label{sec:evolve}

It should be noted that the LUT and its parameters evolve as a function of the energy when standard and non-standard matter effects are taken into consideration.
The case of SI was investigated in~\cite{Zhang:2004hf}. 
We first study the energy dependence of the  full LUT and later investigate the energy dependence of the parameters $b^N_{\alpha\beta}$, $c^N_{\alpha\beta}$ and $\alpha^N_{\alpha\beta}$. 
As it is well known, the area of the LUT is directly related to the Jarlskog factor, $J_{CP}$, and indicates the extent of leptonic CP violation. 
Thus, geometric properties of triangles such as changes in the shape or the area of the LUT are physically relevant. 
In Fig.~\ref{fig:evol}, we illustrate how the LUT gets modified in case of standard and non-standard  matter effects.
Interestingly, for the $\nu_\mu \to \nu_e$ channel (top row of Fig.~\ref{fig:evol}),   and for specific values of NSI couplings: (a) $\varepsilon_{e\mu} = 0.1$, $\varphi_{e\mu} = \pi$, and (b)  $\varepsilon_{e\tau} = 0.1$, $\varphi_{e\tau} = 0$, the triangle tends to shrink to almost a line for $E \sim 2-2.5$ GeV and, beyond this energy range, it starts growing again. 
This is due to the fact that one of the sides ($b^N_{\alpha\beta}$) vanishes around this energy. 
However, no such shrinking is seen in the $\nu_\mu \to \nu_\tau$ channel, at the bottom row of Fig.~\ref{fig:evol}. This observation has interesting connections with the evolution of LUT parameters  as a function of energy and it also has implications on the Jarlskog invariant, as we shall see later.

\begin{figure}[tb!]
\centering
\includegraphics[scale=0.55]{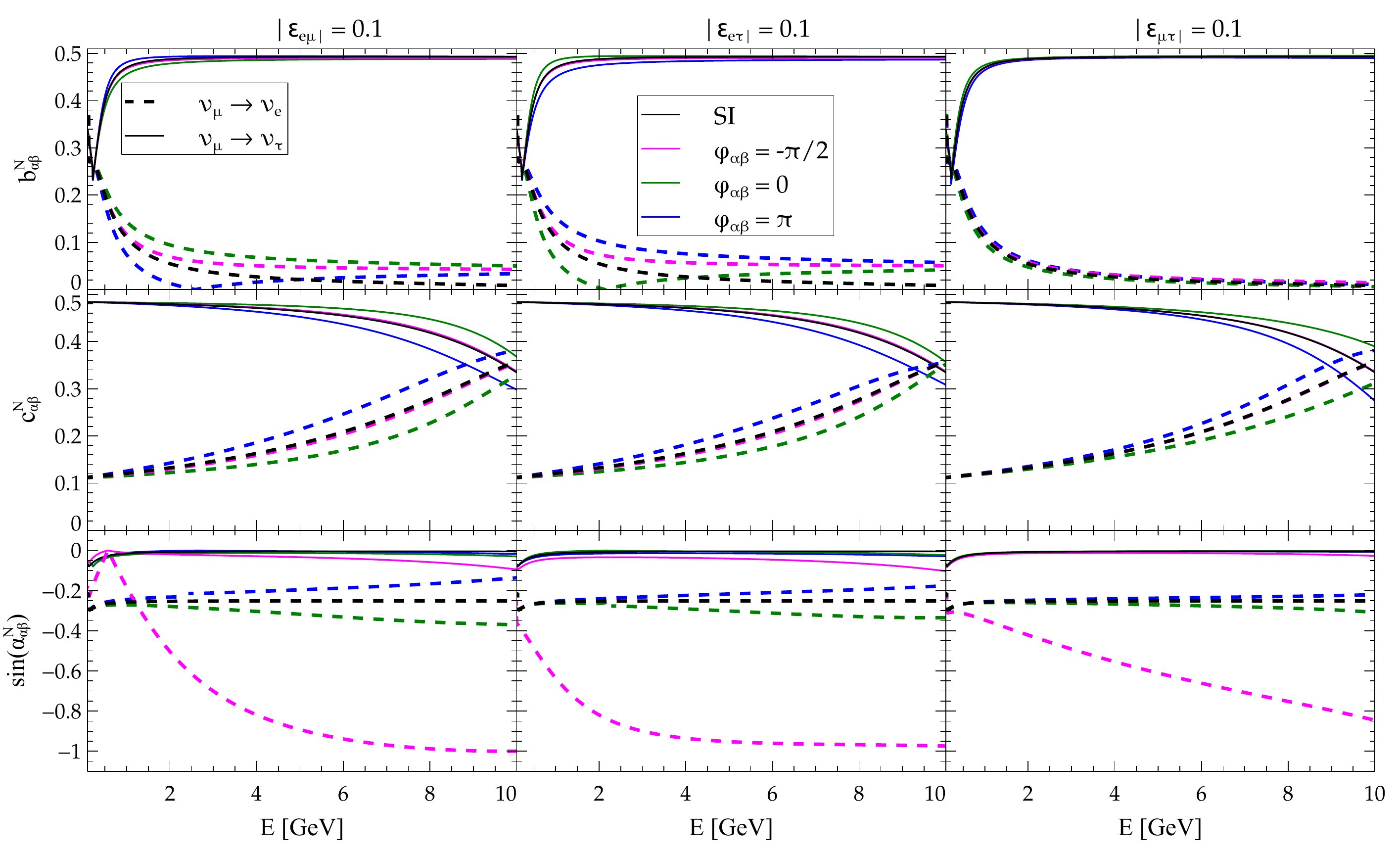}
\caption{\footnotesize{Evolution of the LUT parameters $b^{N}_{\alpha\beta}$, $c^{N}_{\alpha\beta}$ and $\sin\alpha^{N}_{\alpha\beta}$ as a function of the neutrino energy for the $\nu_{\mu} \to \nu_{e}$ and $\nu_{\mu} \to \nu_{\tau}$ channels for the SI and NSI case with different sizes of the couplings and their phases.}}
\label{fig:lut_en_nue_nutau}
\end{figure}

Fig.~\ref{fig:lut_en_nue_nutau} depicts the evolution of the LUT parameters ($b^{N}_{\alpha\beta}, c^{N}_{\alpha\beta}$ and $\sin \alpha^{N}_{\alpha\beta}$) corresponding to the two appearance channels. 
Each column corresponds to the presence of a single NSI coupling   with its value held fixed to the value indicated in the top row.
We have already checked that our LUT probability expression in Eq.~\eqref{eq:prob_nsi} agrees well  with the exact numerical results obtained using GLoBES up to the energies for which the perturbation approach is valid. 
The small disagreement begins to appear only at the higher energy end\footnote{Since the perturbation $A|\varepsilon|$ is proportional to energy, the validity of expansion up to the first order perturbation weakens at higher energy.}.
From Fig.~\ref{fig:lut_en_nue_nutau}, we note that, for the $\nu_{\mu} \to \nu_{\tau}$ channel, the SI and NSI curves are close.  
For the  $\nu_{\mu} \to \nu_{e}$  channel, however, the SI and NSI curves differ more and there are interesting features, namely, 

\begin{itemize}
\item[i)]
for $\varphi_{e\mu} = \pi$  and $E \simeq 2.5$ GeV, we find that $b^N_{\mu e} \to 0$ (top row, left column),
\item[ii)]
for $\varphi_{e\tau} = 0$  and $E \simeq 2.1$ GeV, we find that $b^N_{\mu e} \to 0$ (top row, central column),
\item[iii)]
for $\varphi_{e\mu} = -\pi/2$ and $E \simeq {0.6}$ GeV, $\alpha^N_{\mu e} \to \pi$ (bottom row, left column).
\end{itemize} 

Note that, when $b^{N}_{\alpha\beta}$ vanishes, $\alpha^{N}_{\alpha\beta}$ becomes ill-defined, leading to discontinuities in the curves of $\sin \alpha^{N}_{\alpha\beta}$ at those values.
We have removed these discontinuities while plotting $\sin \alpha^{N}_{\alpha\beta}$ in 
Fig.~\ref{fig:lut_en_nue_nutau}. 
Note that the limits where $b^N_{\alpha\beta}$ vanishes or $\alpha^N_{\alpha\beta}$ goes to $\pi$ imply the triangle shrinking to a line, which means that the NSI parameters conspire to give no CP violation (see also Fig.~\ref{fig:evol}). 
%

\section{Evolution of the Jarlskog Invariant}
\label{sec:jcp}

The Jarlskog factor, $J_{CP}$, is geometric in nature, as it is intimately connected to the area of the LUT, since $J_{CP} = 2 S_{\triangle}$, where $S_{\triangle}$ is the area of the triangle. 
Its importance lies in the fact that it is independent of the phase convention~\cite{Jarlskog:1985ht,PhysRevLett.55.2935,Jarlskog:2004be,Nieves:1987pp,Jenkins:2007ip} and, hence, it is an invariant measurement of CP violation. The impact of different parameterizations on the interpretation of CP violation in neutrino oscillations has been studied in~\cite{Denton:2020igp}. 
For three generations, it turns out that the Jarlskog invariant is uniquely defined (up to a sign) irrespective of the appearance channel. 
Thus, the probability difference  $\Delta P^{CP}_{\alpha \beta} = P ({\nu_\alpha  \to \nu_\beta}) - P  ({\bar \nu_\alpha  \to \bar\nu_\beta}) $ obeys 
\begin{eqnarray}
|\Delta P^{CP}_{\mu e} | = | \Delta P^{CP}_{e \tau} | = | \Delta P^{CP}_{\mu \tau} | \propto J_{CP}~.
\end{eqnarray}
In terms of the LUT parameters, the Jarlskog invariant in presence of NSI can  be written as 
\begin{eqnarray}
J_{CP} &=&   b^{N}_{\alpha\beta} c^{N}_{\alpha\beta} \sin \alpha^{N}_{\alpha\beta}~,  \nonumber
\end{eqnarray} 
where $b^N_{\alpha\beta}$, $c^N_{\alpha\beta}$ and $\alpha^N_{\alpha\beta}$ are the two sides and the angle of the LUT which have been plotted for the  channels $\nu_\mu \to \nu_e$  and  $\nu_\mu \to \nu_\tau$ as a function of energy in Fig.~\ref{fig:lut_en_nue_nutau}.

\begin{figure}[ht!]
\centering
\includegraphics[scale=0.55]{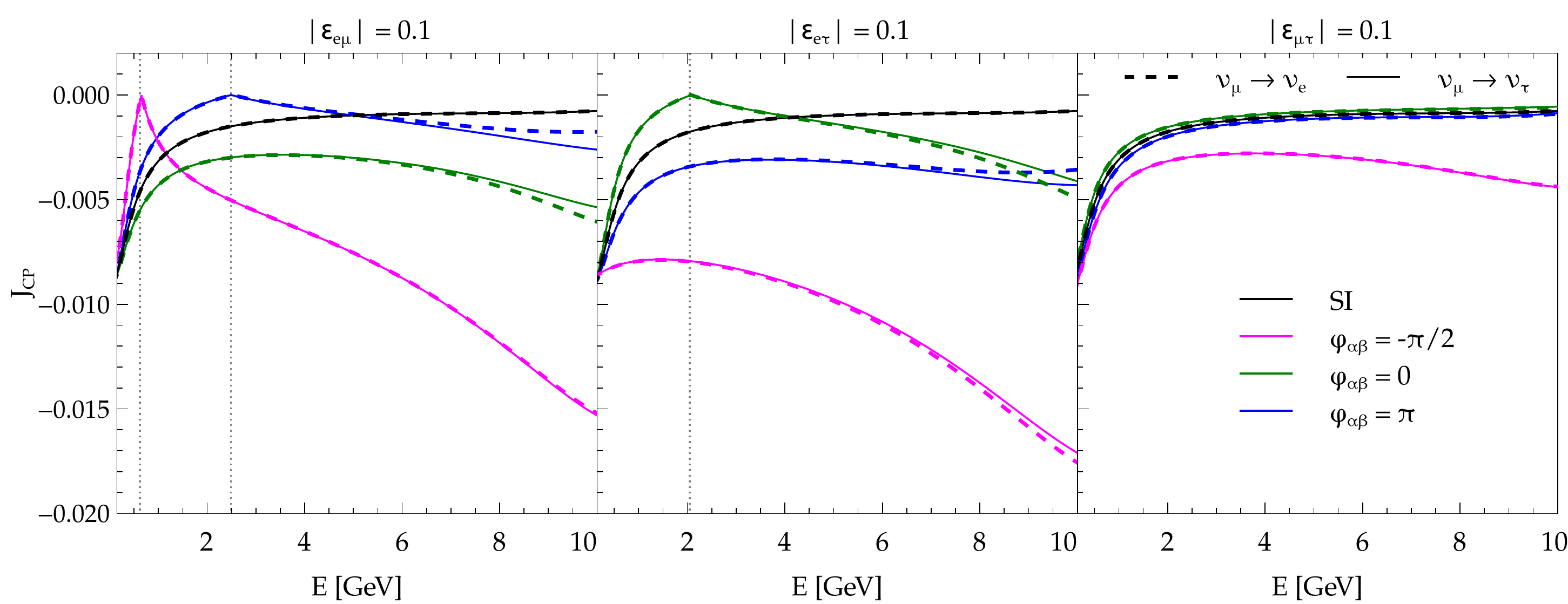}
\caption{Evolution of the Jarlskog invariant as a function of energy in the presence of non-zero NSI couplings ($\eem $, $\eet$, $\emt$) taken one at a time, as indicated in the plot. 
The solid (dashed) lines correspond to the $\nu_{\mu} \to \nu_{\tau}$ ($\nu_{\mu} \to \nu_{e}$) appearance channel for different values of the NSI phases, $\varphi_{\alpha\beta}$.}
\label{fig:lut_jcp}
\end{figure}

In Fig.~\ref{fig:lut_jcp}, we  plot the evolution of $J_{CP}$ as a function of the energy for three different NSI couplings,  $\varepsilon_{e\mu}$, $\varepsilon_{e\tau}$ and $\varepsilon_{\mu\tau}$, along with their respective phases $\varphi_{\alpha\beta}$.
The  Jarlskog invariant computed for the channels $\nu_\mu \to \nu_e$  and  $\nu_\mu \to \nu_\tau$ is shown as dashed and solid curves, respectively. 
The case of SI is depicted in black, while the colours correspond to different choices for the NSI phase $\varphi_{\alpha\beta}$.
At the lower energy limit, the invariant takes its vacuum value given by $J_{CP} \simeq {-0.008}$, which is  obtained from the current best-fit values of neutrino mixing parameters~\cite{deSalas:2020pgw}.
In the presence of neutrino interactions with matter, either standard or non-standard,  the value of $J_{CP}$ rises (with the corresponding decrease in the area of the LUT) or drops (with the corresponding increase in the area of the LUT), depending on the choice of parameters.
In the case of SI, as the energy increases, after an initial  rise, $J_{CP}$ flattens out and saturates to a value close to $-0.001$ for energies above $6$ GeV.

As stated above, since we consider the evolution of three neutrino flavours, there is only one measure of CP violation and the Jarlskog invariant is unique irrespective of the appearance channel considered.
It should be noted that our approximate expression in Eq.~\eqref{eq:prob_nsi} is based on perturbation theory and its validity for the different channels has been discussed in Sec.~\ref{sec:validity}.
As we go to higher energies, above $6$ GeV, the precision starts getting worse for the NSI couplings $\varepsilon_{e\mu}$, $\varepsilon_{e\tau}$ but not as much for $\varepsilon_{\mu\tau}$ {\textcolor{black}{(see Fig.~\ref{fig:prob_precision})}}. This is reflected in the plot of the Jarlskog invariant as well.   
The different  curves in each panel of Fig.~\ref{fig:lut_jcp} correspond to specific choices of the NSI phases, $\varphi_{\alpha \beta} = 0, -\pi/2, \pi$.  It can be seen that the phase dependence is more pronounced for the first (\eem) and second (\eet) panel but not so much for the third panel (\emt). 
Note as well that the SI and NSI curves intersect at a particular value of energy for certain values of the NSI phases:
\begin{itemize}
\item[i)]
at $E \simeq {1.2}$ GeV for $\varphi_{e\mu} = -\pi/2$  and $E \simeq 5.1$ GeV for $\varphi_{e\mu} = \pi$ (first panel),
\item[ii)]   at $E \simeq 4.1$ GeV for $\varphi_{e\tau} = 0$ (second panel).
\end{itemize} 
This means that, around these particular values of the neutrino energy, it will be very difficult to disentangle the origin of CP violation, since the Dirac CP phase, $\delta$, and one of the phases associated to NSI, $\varphi_{\alpha\beta}$, would give rise to the same value of $J_{CP}$. 
%
%
\textcolor{black}{Note, however, that this problem of parameter degeneracy arises only at a specific energy value. Spectral information can help to extract unambiguous information from a given experiment. Thus, if we can obtain information at different energies or at different baselines~\cite{BurguetCastell:2001ez}, we can identify the source of CP violation.}

Interestingly, the value of the Jarlskog invariant vanishes at a particular energy for a certain choice of NSI phases. This is indicated  in Fig.~\ref{fig:lut_jcp} by the vertical dotted lines in the first  and  second panel. 
The vanishing of $J_{CP}$ for a specific values of the NSI phases is in agreement with our observation with regard to the evolution of the LUT in Fig.~\ref{fig:evol} and Fig.~\ref{fig:lut_en_nue_nutau}. 
For $\eemp = \pi$ and $\eetp = 0$, $J_{CP}$ vanishes around $2-2.5$ GeV since one of the sides ($b^{N}_{\alpha\beta}$) vanishes and then the triangle shrinks to a line in Fig.~\ref{fig:evol}. 
Likewise, for $\eemp = -\pi/2$, $J_{CP}$ vanishes around {$0.6$} GeV as the angle, $\alpha^{N}_{\alpha\beta}$ goes to $\pi$, as can be seen in Figs.~\ref{fig:evol} and \ref{fig:lut_en_nue_nutau}.

\section{Conclusion}
\label{sec:conclude}

In the present work, we have formulated an alternative approach based on the LUT to describe neutrino oscillations in the presence of non-standard neutrino interactions with matter.
After briefly reviewing the case of neutrino propagation in vacuum and matter in the presence of standard matter interactions~\cite{He:2016dco}, we have derived the expression for the oscillation probability $P(\nu_{\alpha} \to \nu_{\beta})$ using perturbation theory and cast it in terms of the independent parameters of the LUT, labeled as $b^N_{\alpha\beta}, c^N_{\alpha\beta}$ and $\sin\alpha^N_{\alpha\beta}$ in Eq.~\eqref{eq:prob_nsi}. 
Even though NSI introduces additional  parameters, the form of the oscillation probability retains the same structure as in vacuum or standard matter interactions with the appropriate redefinitions of the sides and angles of the LUT. 
Obtaining Eq.~\eqref{eq:prob_nsi} and thereby establishing the robustness of the expression of the neutrino oscillation probability in terms of LUT parameters in the presence of NSI constitutes the key result of this work. Thus, the invariant form of the probability expression in the presence of a new physics scenario\footnote{The form invariance is retained as long as the mixing matrix is unitary, and it is expected to break down for the case of additional sterile neutrinos as the mixing matrix becomes non-unitary.} facilitates a neat geometric view of neutrino oscillations in terms of LUT as depicted in Fig.~\ref{fig:uni_tri}.

We examined the validity of the analytic result obtained in Sec.~\ref{sec:validity} and  discussed the role played by the NSI couplings in the energy evolution of the LUT parameters as well as the triangle in Sec.~\ref{sec:evolve}. We find that, for some specific choices of NSI terms and energy, the triangle shrinks to a line. Finally, we studied the   evolution of the Jarlskog invariant as a function of energy for the NSI case and compare it with the SI case in Sec.~\ref{sec:jcp}.
The geometric  approach based on LUT allows us to express the oscillation probabilities for a given pair of neutrino flavours in terms of only three degrees of freedom which are related to the geometric properties,  two sides and one angle, of the unitarity triangle. 
Moreover, the LUT parameters are invariant under rephasing tranformations and independent of the parametrization  adopted. These are the main advantages of this approach.

\section*{Acknowledgements} 
{\hlpm{It is a pleasure to thank Peter Denton for valuable comments. 
}}
MM is supported by IBS under the project code IBS-R018-D1.  PM is supported by funding from  University Grants Commission under  UPE II at JNU and  Department of Science and Technology under   DST-PURSE at JNU. The use of HPC cluster at SPS, JNU funded by DST - FIST is acknowledged. PM is also supported by the European Union's Horizon 2020 research and innovation programme under the Marie Skodowska-Curie grant agreement No 690575 and 674896. 
CAT is supported by the research grant ``The Dark Universe: A Synergic Multimessenger Approach'' number 2017X7X85K under the program ``PRIN 2017'' funded by the Ministero dell'Istruzione, Universit\`a e della Ricerca (MIUR).
MT is supported by the Spanish grants FPA2017-85216-P (AEI/FEDER, UE),PROMETEO/2018/165 (Generalitat Valenciana) and the Spanish Red Consolider MultiDark FPA2017-90566-REDC.


\bibliography{bibliography}

\end{document}